\documentclass[epjST]{svjour}
\usepackage{graphics}

\newcommand{\C}[1]{{\mathcal{#1}}}
\newcommand{\ket}{\rangle}
\newcommand{\bra}{\langle}

\newcommand{\spc}{|}

\begin{document}
\title{Coherent control of quantum transport: modulation-enhanced phase detection and band spectroscopy}
\author{Marco G. Tarallo\inst{1}\fnmsep\thanks{\email{tarallo@lens.unifi.it}} \and Nicola Poli\inst{1}, F.--Y. Wang\inst{1} \and Guglielmo M. Tino\inst{1} }
\institute{Laboratorio Europeo di Spettroscopia Nonlineare -- LENS, Universit\`a di Firenze, Via Nello Carrara 1, 50019, Italy. }
\abstract{
Amplitude modulation of a tilted optical lattice can be used to steer the quantum transport of matter wave packets in a very flexible way. This allows the experimental study of the phase sensitivity in a multimode interferometer based on delocalization-enhanced Bloch oscillations and to probe the band structure modified by a constant force.
} 
\maketitle
\section{Introduction}
\label{intro}

Cold atoms trapped in an optical lattice represents a perfect tool to implement and study different phenomena predicted by quantum mechanics in many fields such as condensed matter and solid state physics~\cite{Bloch} or quantum information processing~\cite{Bloch2} and to simulate gauge field theories~\cite{gauge}. In particular, the atomic motion in the optical lattice can be modified and driven by the external forces, like the gravitational force or a static electromagnetic field, or by perturbing the optical lattice field itself, like the presence of a phase or amplitude modulation or the scattered field associated to the presence of rough surfaces along the optical path. The interplay between the external driving force and optical lattice field modulation has been recently studied extensively with the objective of the realization of a precise force sensor~\cite{myself}. 

In this paper, we describe two particular phenomena concerning a driven optical lattice. The first considers the multimode interference between neighborhood quantum states populated in a optical lattice. The second phenomenon considers the experimental investigation of the multiband structure of the energy spectrum and the associated transport.

The paper is organized as follows: in Sect.~\ref{sec:1}, we briefly recall the main features of the dynamical system, i.e. a tilted optical lattice subjected to amplitude modulation of the intensity. Section~\ref{sec:2} deals with the analysis of our multimode atom interferometer with a discussion about the interference peak width. Finally, Sect.\ref{sec:3} shows the effect of interband resonant transitions induced by the amplitude modulation and a possible application for the study of the band structure and calibration of the trap depth.

\section{Quantum dynamics in driven optical lattices}
\label{sec:1}

We experimentally study the dynamics of non-interacting ultracold atoms of mass $M$ in an accelerated one-dimensional optical lattice under the effect of a periodic driving of its amplitude. The effective Hamiltonian describing the system is

\begin{equation}\label{eq:startH}
\C{H}(z,p,t) =  \frac{p^2}{2 M}+\frac{U_0}{2}\cos(2k_Lz)[1+\alpha \sin(\omega t)] +Fz\,,
\end{equation}
where a periodic potential $(U_0/2)\cos(2k_Lz)\equiv U(z)$ is originated by the interference pattern of two counterpropagating laser beams with wavelength $\lambda_L = 2\pi/k_L$, $U_0$ is the depth of the lattice potential which is modulated with amplitude $\alpha$. This system has been extensively studied both theoretically and experimentally. In the absence of the amplitude modulation (AM), the external constant force $F$ breaks the translational symmetry due to the presence of the periodic potential $U(z)$. This results in the formation of the so-called Wannier-Stark (WS) ladders~\cite{Niu96}, where the eigenstates $\spc\Psi_{n,j}\ket$ are localized quasi-stationary states constructed from the energy band of index $j$ and centered on the lattice site of index $n$ \cite{Gluck:2002p767}. The quasi-energies of the $j$th ladder $\C{E}_j$ are centered on the average energy value of the corresponding band $\C{E}_j=(2k_L)^{-1}\int_{-k_L}^{k_L}dk\ E_j(k)$, while WS state energies are mismatched at different sites by a constant step $Fd=\hbar\omega_B$, where $d=\lambda_L/2$ is the lattice period and $\omega_B$ is the Bloch frequency. 

In this system, quantum transport occurs as interference over macroscopic regions of phase space and tunneling between different ladders. In fact, because of the phase imprint $\omega_Bt$ which occurs from site to site~\cite{Roati04}, a wavepacket prepared in a superposition of WS states undergoes oscillations with angular frequency $\omega_B$ both in momentum and real space and are called Bloch oscillations (BO). The BO phenomenon is very well known in atomic physics since the seminal work done by BenDahan et al.~\cite{BenDahan} and can be interpreted as a result of coherent interference among the WS states which compose the atomic wavepacket. On the other hand, it is possible that atoms subjected to BOs may undergo interband tunneling, known as  Landau-Zener tunneling (LZ)~\cite{Holt2000}.

In this scenario, the periodic driving $\alpha (U_0/2)\cos(2k_Lz)\sin(\omega t)$ represents a flexible tool to manipulate quantum transport by inducing resonant transitions between WS states. In the case of atoms lying on the lowest energy band, resonant tunneling occurs when the atoms absorb or emit energy quanta $\hbar\omega$ which satisfy the condition

\begin{equation}\label{eq:WSladders}
\hbar\omega+\C{E}_0-\C{E}_j-\ell\hbar\omega_B = 0\, ,
\end{equation}
where we consider the integer $\ell$ to be the lattice site difference between the resonantly coupled WS states. In the following, we will discuss two AM regimes: for $\hbar\omega \ll \C{E}_1-\C{E}_0$ intraband resonant tunneling takes place~\cite{Ivanov}, while for $\hbar\omega\geq\C{E}_1-\C{E}_0$, it is possible to induce resonant transitions between WS ladders~\cite{Wilkinson96,Gluck99}.


\subsection{Intraband resonant tunneling: delocalization-enhanced Bloch oscillations}

We first consider a system of an ultracold atomic gas with sub-recoil temperature under the effect of a weak constant force $F$ and AM frequencies $\omega$ smaller than the first band gap. In this regime LZ tunneling is fully negligible, thus we drop the band index $j$, since under this assumption the atomic dynamics remains confined in the same initial lattice band at all times. Coherent delocalization of matter waves by means of intraband resonant tunneling is established when near-resonant modulation is applied, i.e. $\omega \simeq\ell\,\omega_B$. 

It is possible to demonstrate that the Hamiltonian~(\ref{eq:startH}) can be rewritten as

\begin{equation}\label{eq:WS_AMham} 
\C{H}'_{\mathrm{AM}} =\sum_{n=-\infty}^{+\infty}\bigg[n\frac{\hbar\delta}{\ell} \spc
\Psi_{n}\ket\bra\Psi_{n}\spc
+\left(i\frac{\C{J}_\ell}{2}\spc \Psi_{n+\ell}\ket \bra \Psi_{n}\spc +\mathrm{h.c.}\hspace{-1pt}\right)\bigg]\, ,
\end{equation}
where $\delta=\omega-\ell\omega_B$ is the detuning and $\C{J}_\ell=(\alpha U_0/2)\bra \Psi_{n+\ell}\spc\cos(2k_Lz)\spc\Psi_{n}\ket$ represents the tunneling rate. In this condition, the system exhibits a strict analogy with the Hamiltonian of a static lattice in the presence of an effective homogenous force of magnitude $F_\delta=\hbar\delta/(\ell d)$. 

In the case of small detunings ($\delta\neq 0$), the transport dynamics is characterized by Bloch oscillations with time period $2\pi/\delta$, which was first observed through macroscopic oscillations of atomic wavepackets \cite{Alberti:2009p45}, and later studied with non-interacting Bose-Einstein condensates (BECs) \cite{Haller:2010hx}. In the other case with $\delta=0$, the system is invariant under discrete translations, and coherent delocalization occurs, with the atomic wavepackets spreading ballistically in time. This results in a broadening of the width of the atomic spatial distribution according to $\sigma(t)\approx \C{J}_\ell \ell d\, t/\hbar $~\cite{Ivanov}.

Hence, this system is a perfect tool  to engineer matter-wave transport over macroscopic distances in lattice potentials with high relevance to atom interferometry.



\subsection{Interband resonant tunneling: Wannier-Stark spectroscopy}

When $\hbar\omega\geq\C{E}_0-\C{E}_j$, AM can drive resonant interband transitions~\cite{Friebel,Stoeferle}. In particular, in the dynamical system depicted in Eq.\ref{eq:startH}, it is possible to induce resonant transitions between WS ladders~\cite{Wilkinson96,Gluck99}. In this regime, it is not possible to neglect the non-zero lifetime of the excited ladders due to the LZ tunneling effect, which becomes the governing quantum transport phenomenon. Then the measurement of the number of remaining atoms after modulation of the trap depth (which is proportional to the survival probability $P(\omega)$) provides a direct observation of the WS ladders spectrum. If the AM is applied for a time $t_{\mathrm{mod}}$, the fraction of remaining atoms inside the optical lattice is given by

\begin{equation}\label{eq:eq3}
P(\omega)\simeq\exp\left[-\left(\frac{\alpha U_0}{2\hbar}\right)^2\,t_{\mathrm{mod}}\, \sum_{j,\ell}\frac{\C{C}_{j,\ell}^2\ \Gamma_j}{\Delta_{0,j}(\ell)^2+\Gamma_j^2}\right]\, ,
\end{equation}
where the coefficients $\C{C}_{j,\ell}=\bra \Psi_{j,n+\ell}\spc \cos(2k_Lz)\spc \Psi_{0,n}\ket$ are the real-valued overlap integrals between resonantly-coupled WS states (only transitions from the first ladder ``0'' are considered), $\Delta_{0,j}(\ell)=\omega-(\C{E}_j-\C{E}_0)/\hbar-\ell\omega_B$ is the AM frequency detuning. 

In the case of a strong tilt, i.e. $Fd \simeq E_r$, it is possible to resolve each single resonance occurring at $\ell\omega_B+(\C{E}_0-\C{E}_j)/\hbar$ and thus measure $\omega_B$ with high precision~\cite{madison2}. In the opposite regime, i.e. $Fd \ll E_r$, the depletion spectrum will present a comb of resonances around the WS ladder central frequencies $\omega_{0,j}=(\C{E}_j-\C{E}_0)/\hbar$. In this case, the external force represents only a small perturbation to the band structure and interband transitions may probe the band energy dispersion~\cite{Heinze}.




%

\section{A multimode atom interferometer: DEBO phase sensitivity}
\label{sec:2}

Let us first consider our system consisting of a tilted optical lattice as an example of multimode matter-wave interferometer. As previously stated in Sect.~\ref{sec:1}, a wavepacket which occupies $N$ lattice sites can be viewed as a coherent superposition of $N$-WS states evolving with a Bloch phase $\theta_B(t) = n\,\omega_B t=\frac{\pi}{k_L}k(t)$. In the case of a BEC, the number of coherent WS states is $N\simeq \sigma_{\mathrm{BEC}}/d$, which can interfere in time-of-flight (TOF). By increasing the number of populated lattice sites $N$, it is predicted an enhancement of the Bloch phase determination inferred by a least-squares fit as~\cite{Piazza}

\begin{equation}\label{eq:5}
\Delta\theta_B \simeq \sqrt{\frac{1}{mN_{\mathrm{at}}}\times\frac{1}{N^2}}\,,
\end{equation}
where $m$ is the number of repetitions of the experiment and $N_{at}$ is the number of atoms. The latter formula has an evident confirmation if we consider that the width of the interference peak $\Delta k\propto 1/N$. The enhancement of the Bloch phase sensitivity in an incoherent ensemble of ultracold atoms, which lets high precision measurements of gravity by means of the delocalization-enhanced Bloch oscillation (DEBO) technique~\cite{Poli2011,myself}, is also based on the increase of the number of populated lattice sites $N$. Here we analyze our previous experimental data and perform numerical calculations about DEBO multimode interferometer in order to quantitatively validate Eq.\ref{eq:5} as predicted by Ref.~\cite{Piazza}.

The DEBO technique consists of broadening of each atomic wavefunction by means of a resonant AM burst for a time interval $t_{\mathrm{mod}}$, so that the wavefunction is delocalized over $N\simeq \C{J}_1t_{\mathrm{mod}}/\hbar$ lattice sites. After evolving in the lattice, the atoms are suddenly released to produce a narrow interference pattern which occurs at the transient time $\tau_{\mathrm{trans}}=MNd^2/\hbar$~\cite{myself}.

\begin{figure}[t]
\centering
\resizebox{0.70\columnwidth}{!}{%
  \includegraphics{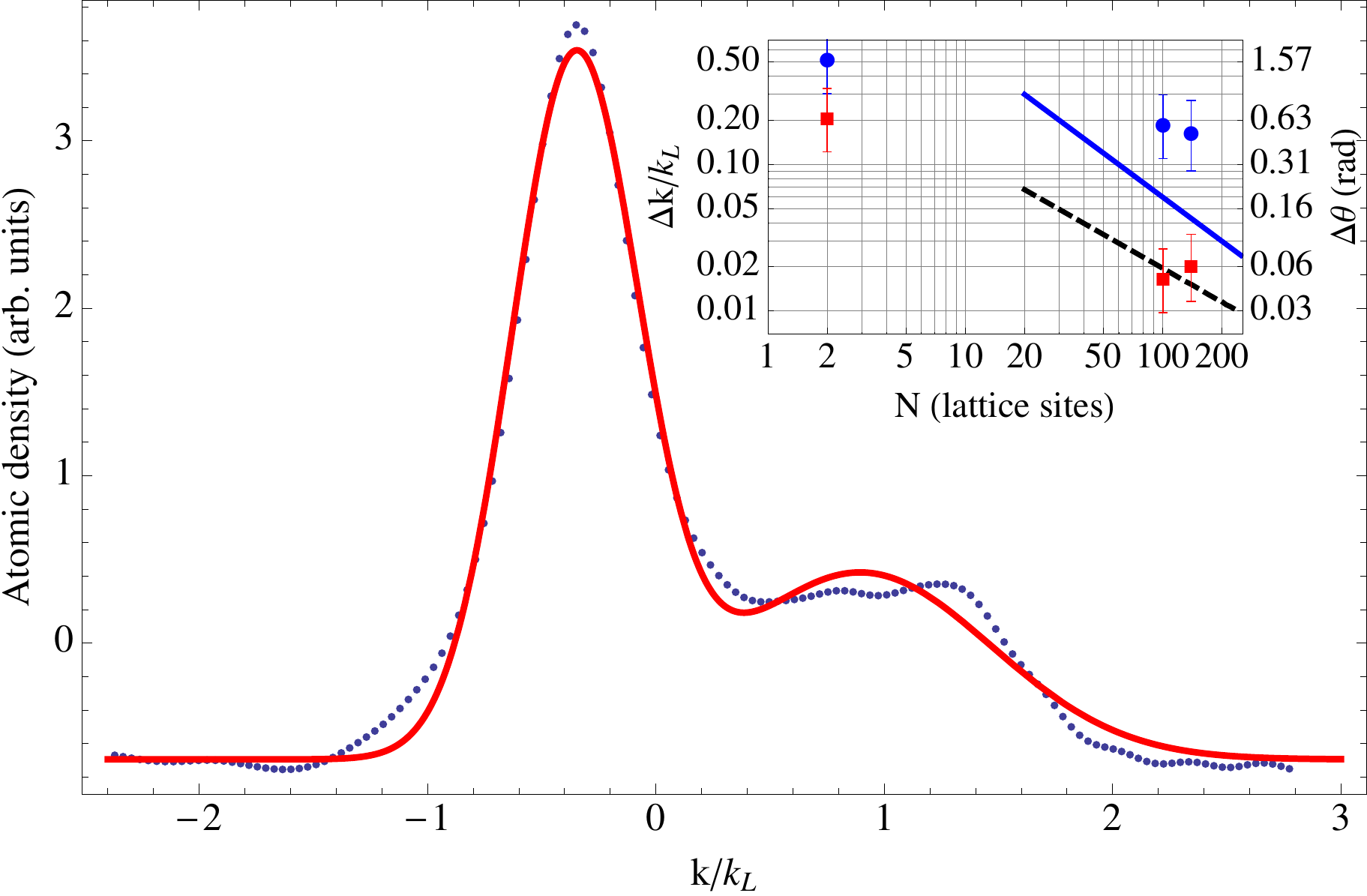} }
\caption{Experimental detection of the DEBO transient interference: from the atomic density distribution in TOF a least-squares fit extracts the momentum value $k(t)$ and thus the Bloch phase with an error of $\Delta\theta_B$. The inset shows numerical calculations (lines) of DEBO effect compared with typical experimental results (points). The dashed line is the $1/e^2$ width of the wavefunction after TOF at t = $\tau_{\mathrm{trans}}$; the solid line is the rms width of the atomic cloud which broadens the smallest detectable $\Delta k$; circles and squares represent the peak width and the average Bloch phase error as a function of the number of occupied lattice sites $N$, respectively.}
\label{fig:3}       
\end{figure}
We realize the multimode interferometer as follows: we load an ensemble of $5\cdot10^4$ ultracold $^{88}$Sr atoms at 1 $\mu$K temperature into a vertical optical lattice with $\lambda_L$ = 532 nm accelerated by the gravity force. Here the temperature reduces up to $T \sim 0.6\,\mu$K by evaporation, so that the atoms fill only the lowest band. We then apply a servo-controlled AM burst with $\omega=\omega_B$ delocalizing the atom's wavefunction over $N$ = 100 and 140 lattice sites. Figure~\ref{fig:3} shows a typical density profile after TOF detection of the DEBO interference peak. Here a narrow peak emerges from a broad distribution which reflects the WS momentum density which is flat over the whole Brillouin zone. Typical results for our experiment are $\Delta k$ = $0.18\, k_L$ (blue circles) which corresponds to an error on the momentum $\sigma_k = 1.6\cdot10^{-2}\,k_L$ and thus on the Bloch phase $\Delta\theta_B =5\cdot10^{-2}$ rad. We compare this result with the typical sensitivity on the Bloch phase of a $^{88}$Sr gas without the use of the DEBO technique, where $N\sim2$ is given by the thermal de Broglie wavelength. In the inset of Fig.~\ref{fig:3} we plot the Bloch phase uncertainty (red squares) by means of DEBO technique for different values of N, and we find that the error on the phase decreases as $\Delta\theta_B \propto N^{-0.7}$.

We investigated the DEBO phase sensitivity also by numerical calculations. We calculated the single wavefunction at the end of the DEBO sequence for different values of $N$ and then we performed a Gaussian least-squares fit to estimate the resulting $\Delta k$. The dashed line in Fig.~\ref{fig:3} shows that the dependence on the number of populated lattice sites is $\Delta k \propto N^{-3/4}$. However, this curve is still below the minimum detectable atomic cloud width which correspond to the initial cloud size $\sigma_z$ (solid line): by increasing the amplitude of the wavefunction, and thus of the interference time $\tau_{\mathrm{trans}}$, the relative weight of $\sigma_z$ decreases as $1/N$. 

Comparing the numerical calculations with the experimental data, we see that the experimental $\Delta k$ is about a factor of 3 higher than the $\sigma_z$ limit, mainly due to the interference pattern visibility $\C{V}\equiv (OD(\mathrm{max})-OD(\mathrm{min}))/OD(\mathrm{max})+OD(\mathrm{min}))\sim 0.6$~\cite{Gerbier}, where $OD$ is the optical density of the atomic sample during absorption imaging. Regarding the Bloch phase error, we can approximate $\Delta\theta_B\simeq \Delta z_0\Delta k/\C{V}$, where $\Delta z_0\sim 5$\% is the relative error on the initial atomic cloud position, which is the major source of uncertainty on the determination of the interference peak.

\section{Amplitude-modulation induced interband transitions and transport analysis}
\label{sec:3}

We have experimentally investigated AM-induced interband transitions in a vertical optical lattice on the same experimental setup as described in Sect.~\ref{sec:2}. In this case, we performed \textit{in-situ} WS ladders spectroscopy by applying an AM perturbation to the depth of the lattice potential with $\alpha$ = 5.3 \% for a modulation time $t_{\mathrm{mod}}$ = 250 ms. We used the AM driving as probing field, scanning its frequency between 15 and 80 kHz and then counting the number of atoms remaining in the lattice by absorption imaging. The depth of the lattice was $U_0$ = 2.5 $E_r$, where $U_0$ has been calibrated as described in Ref.\cite{myself} and the recoil energy $E_r$= 8 kHz $\times h$. Given these numbers, we estimate that the interband transitions are centered around $\omega_{0,1}$ = 20 kHz and $\omega_{0,2}$ = 51 kHz. 

\begin{figure}[t]
\centering
\resizebox{0.70\columnwidth}{!}{%
  \includegraphics{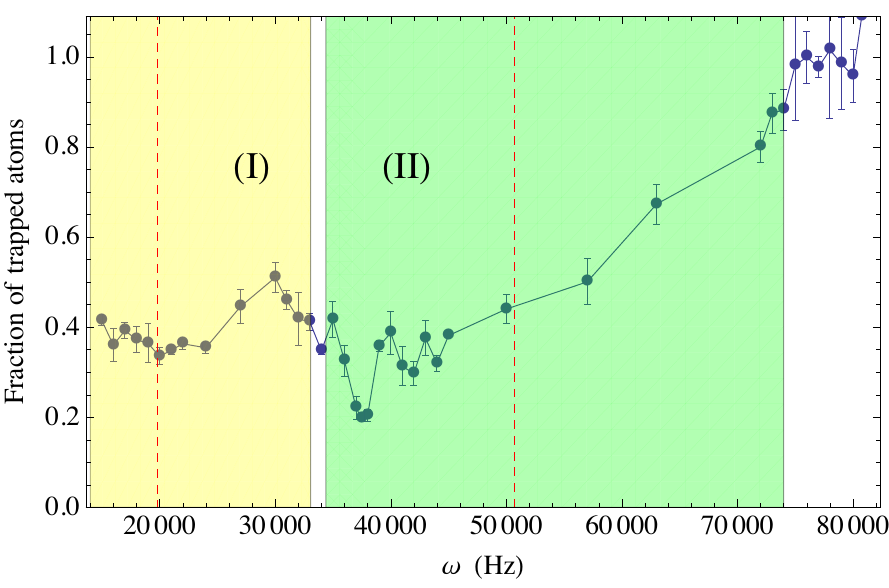} }
\caption{Experimental spectrum of the survival probability as a function of the modulation frequency $\omega$. The shaded regions correspond to the frequencies at which interband transitions with $\Delta k=0$ are expected from the lowest band. The (red) dashed lines represent the resonances $\omega_{0,1}(0)$ = 20 kHz and $\omega_{0,2}(0)$ = 51 kHz.}
\label{fig:1}       
\end{figure}
Figure~\ref{fig:1} shows the fraction of remaining atoms as a function of the modulation frequency. Here we can notice two broad depletion features centered at about 20 kHz and 45 kHz, while a deep and narrow resonance occurs at 37 kHz.

We compare the observed spectrum with the expected positions of the interband transitions by superimposing two colored regions on it, which correspond to the expected frequencies at which transitions to the first and second excited bands (``I''  and ``II'' in Fig.\ref{fig:1}, respectively) should occur with $\Delta k =0$. In particular, in the frequency range between 15 and 30 kHz we observe a symmetric depletion centered around $\omega=\omega_{0,1}$ which approximatively corresponds to the first excited band. At about $\omega$ = 30 kHz there is a local maximum of the fraction of remaining atoms, while the expected gap should appear at 34 kHz. For $\omega >$ 34 kHz, there is a second wide depletion region which corresponds to transitions to the second excited band. However, in this case we notice a marked asymmetry with the minimum $P(\omega)$ region close to the gap frequency. Finally, the largest resonance peak occurs at 37 kHz which is about $2\omega_{0,1}$, with a linewidth of about 2 kHz.

Regarding the interband transitions to the first excited band, in our regime of a weak tilt on a shallow optical lattice ($\omega_B$ = 574 Hz $\ll\, U_0$), the expected LZ linewidth is $\Gamma_1$ = 360 Hz, which is comparable to the Bloch frequency. Hence, it is not possible to resolve the ladder resonances and the presence of several accessible vertical states is responsible of the broad resonance feature. If we look at the interband transitions to second excited band, we expect that its WS states are fully delocalized so that the ladder is no longer discrete. Then the deepest resonance should be due to a parametric excitation mechanism with one of those states~\cite{Friebel}. We also notice that the asymmetry in the excited fraction of atoms at the beginning of the band ($k \simeq 0$) is also present both in the numerical calculation carried on Ref.\cite{Gluck99} (although the calculation is performed for a phase-modulated lattice) and in the experimental data reported in~\cite{Heinze} (although this experiment does not employ a tilted optical lattice).

\begin{figure}
\centering
\resizebox{0.70\columnwidth}{!}{%
  \includegraphics{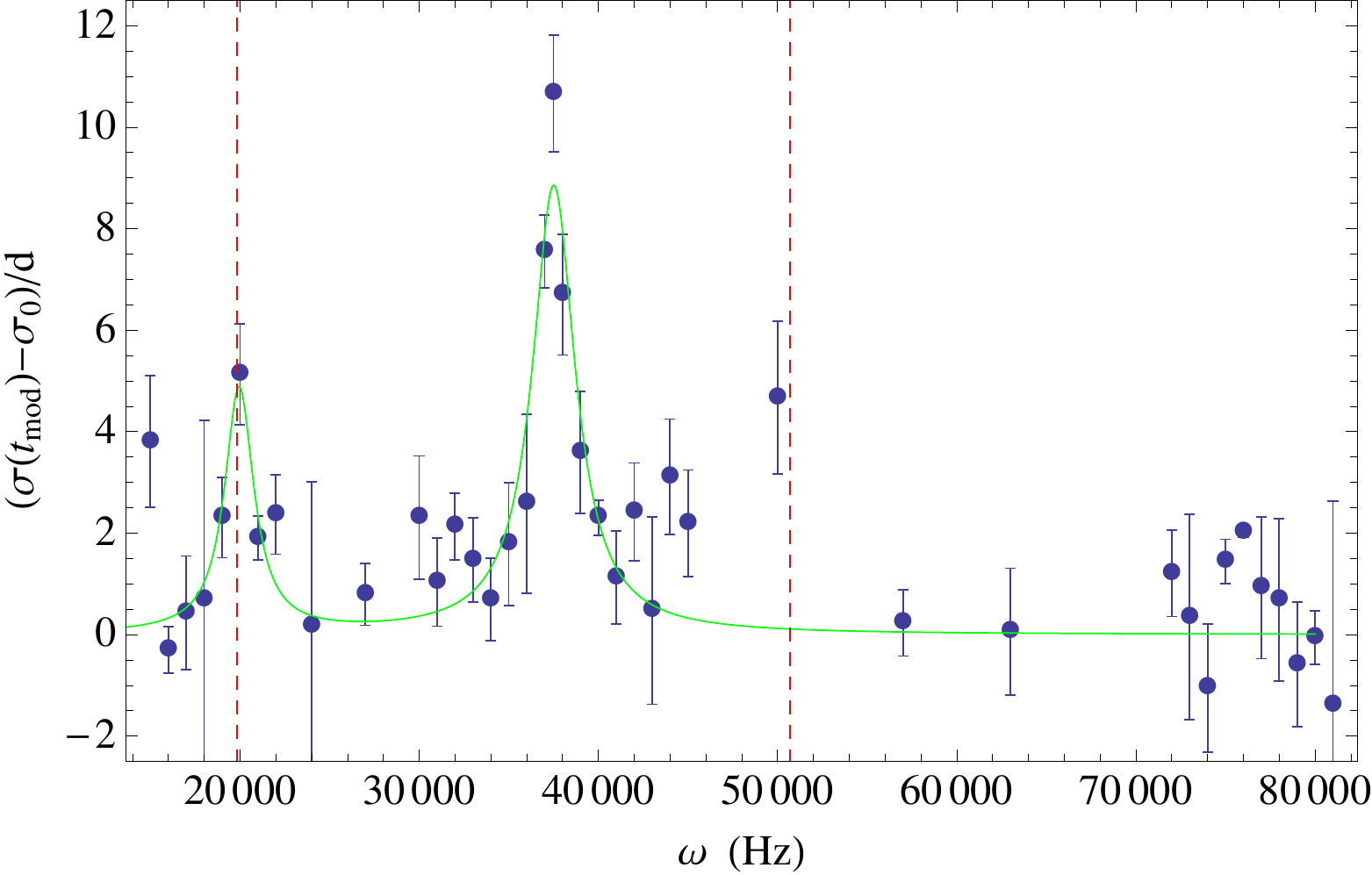} }
\caption{Experimental study of transport resonances during AM induced interband transitions. Each data point represents the difference between the \textit{in-situ} atomic cloud width after AM perturbation and the unperturbed one. As reported in Fig.\ref{fig:1}, the (red) dashed lines correspond to the interband expected resonances with $\ell=0$. The (green) line is a double-peak Lorentzian function best fit of the spatial broadening spectrum.}
\label{fig:2}       
\end{figure}

Since we employ the system for high sensitivity quantum transport measurements, we also observed the corresponding \textit{in-situ} broadening of the spatial atomic distribution as a function of the applied AM frequency. The result is shown in Fig.\ref{fig:2}. Here it is possible to observe two transport resonances at 19.9(3) kHz and 37.5(3) kHz. The amplitudes of these resonances range only between 5 and 9 lattice sites $d$. By observing the fluctuations of the rms width of the atomic cloud $\sigma_0$,  we find that they follow a Gaussian statistics with standard deviation $\Delta\sigma_0 < 2\, d$. Then the two broadening amplitudes are both larger than $2\Delta\sigma_0$.


The observed transport resonance widths are of the order of 2 kHz, which is larger than the natural LZ linewidth $\Gamma_1$. However, in the case of the first resonance, this can be the result of the merging of several transitions with $|\ell|>0$. Assuming that this resonance corresponds to $\omega_{0,1}$, we can calculate the corresponding trap depth value. We find that for $\omega_{0,1}$ = 19.9(3) kHz, the lattice depth is 2.54(4) $E_r$, which is in agreement with the calibration method of Ref.\cite{myself} and has the same relative precision.

\section{Conclusions and outlook}

In this paper we reported on two applications of quantum transport in a driven optical lattice, the first concerning the use of AM at the Bloch frequency to study the sensitivity of a multimode atom interferometer, the second concerning the AM excitation of interband transitions. 

We compared the predicted $\Delta\theta_B \propto N^{-1}$ reduction of the phase uncertainty~\cite{Piazza} with the performances of our system of incoherent ultracold Sr gas by means of the DEBO technique. In our case, we observed a valuable increase of sensitivity which scales nearly as $N^{-0.7}$. This scaling law is confirmed by numerical calculations, where the interference peak width $\Delta k\propto N^{-3/4}$. The reduced exponent in the DEBO case can be due to the non-uniform probability of being at a particular lattice site.

We then studied the response of the atomic system at modulation frequencies larger than the frequency separation between the lowest and the excited bands. We observed a clear frequency-dependent depletion of the lattice population, with the emergence of a gap between the first and the second band. A similar experiment was recently carried out to measure local gravity~\cite{Tack}, where the WS spectroscopy was performed by means of Raman transitions instead of AM of the lattice potential, but in this case the band structure was not resolved. We also observed a weak broadening of the atomic spatial distribution at two different modulation frequencies. The first of these corresponds to the center of the first excited band. Using this transport resonance, we calibrated the lattice depth with 5\% relative uncertainty. Further numerical investigation could provide a better model for either the depletion spectrum and the presence of transport resonances which can become a useful tool to perform band spectroscopy, as done in the case of degenerate Fermi gases~\cite{Heinze}.

\paragraph{Acknowledgments} 
We acknowledge INFN, POR-FSE 2007-2013, iSense (EU--FP7) and LENS (Contract No.RII3 CT 2003 506350) for funding. We thank M.L. Chiofalo and A. Alberti for the theoretical model, S. Chauduri and F. Piazza for critical reading of the manuscript and useful discussions.


\end{document}